\documentclass[twocolumn]{el-author}

\usepackage{graphicx}
\usepackage{subfig}
\usepackage{caption}
\usepackage{float}
\newcommand{\ua}{\uparrow}
\newcommand{\nc}{\newcommand}
\nc{\da}{\downarrow} \nc{\hc}{\hat{c}} \nc{\hS}{\hat{S}}
\nc{\bra}{\langle} \nc{\ket}{\rangle} \nc{\eq}{equation (\ref}
\nc{\h}{\hat} \nc{\hT}{\h{T}}\nc{\be}{\begin{eqnarray}}
\nc{\ee}{\end{eqnarray}}\nc{\rd}{\textrm{d}}\nc{\e}{eqnarray}\nc{\hR}{\hat{R}}\nc{\Tr}{\mathrm{Tr}}
\nc{\tS}{\tilde{S}}\nc{\tr}{\mathrm{tr}}\nc{\8}{\infty}\nc{\lgs}{\bra\ua,\phi|}\nc{\rgs}{|\ua,\phi\ket}
\nc{\hU}{\hat{U}}\nc{\lfs}{\bra\phi|}\nc{\rfs}{|\phi\ket}\nc{\hZ}{\hat{Z}}\nc{\hd}{\hat{d}}\nc{\mD}{\mathcal{D}}
\nc{\bd}{\bar{d}}\nc{\bc}{\bar{c}}\nc{\mc}{\mathcal}\nc{\ea}{eqnarray}\nc{\mG}{\mathcal{G}}\nc{\bce}{\begin{center}}
\nc{\ece}{\end{center}}

\usepackage{cite}
\usepackage{amsmath}

\makeatother

\begin{document}

\title{Fluorescent Troffer-powered Internet of Things: An Experimental Study of Electric-field Energy Harvesting}

\author{O.~Cetinkaya and Ozgur~B.~Akan}

\abstract{A totally new energy harvesting architecture that exploits ambient electric-field (E-field) emitting from fluorescent light fixtures is presented. A copper plate, $50$x$50$cm in size, is placed in between the ambient field to extract energy by capacitive coupling. A low voltage prototype is designed, structured and tested on a conventional ceiling-type 4-light fluorescent troffer operating in $50$Hz $220$V AC power grid. It is examined that the harvester is able to collect roughly $1.25$J of energy in $25$min when a $0.1$F of super-capacitor is employed. The equivalent circuit and the physical model of the proposed harvesting paradigm are provided, and the attainable power is evaluated in both theoretical and experimental manner. The scavenged energy is planned to be utilized for building battery-less Internet of Things (IoT) networks that are obliged to sense environmental parameters, analyze the gathered data, and remotely inform a higher authority within predefined periods.}

\maketitle

\section{Introduction}

Thanks to the advances in ultra-low power transceiver technology, plenty of source such as light, heat, motion and electromagnetic waves/fields have become more viable to build wireless sensor networks (WSNs) which are free from battery constraints \cite{ozgur}. Although the conducted researches theoretically and experimentally revealed the capabilities of these techniques in providing adequate and stable power, rising and diversified needs of today's communication architectures such as IoT, require more enhanced, robust and durable power provision systems. In this regard, electric field energy harvesting stands as the most promising candidate with the characteristics of ambient variable in-dependency, sufficient power rating, low complexity, and excellent energy continuity \cite{Linear, low3, low2, EFEH, Moghe, patent}. 

When an office-like commercial environment is envisioned, illumination can be referred as one of the most crucial systems that needs to be operated perpetually, mostly due to security issues. Regarding this fact, and the strong E-field gradient in the vicinity of fluorescent light fixtures a reference procedure has been represented in \cite{Linear} as the very first approach.
This model is able to provide $200\mu$W of DC power with respect to the designated configuration; however, the setup is bulk, hard to employ, and affects light propagation. Regarding these issues we propose a new architecture that enables ease of implementation, less complex circuitry, and highly increased efficiency as allowing self sufficient IoT networks to be built for online condition monitoring.

\section{E-field Energy Harvesting}

According to the basics of electrostatics, any conductive material energized at some voltage level emits a radial electric field. In AC, this time varying field results in a displacement current $I_{D}$ which dispatches the E-Field induced electric charges to be collected in a storing element $C_{s}$. Since the stored energy is harnessed from the surrounding field, this method can be referred as Electric Field Energy Harvesting (EFEH) \cite{low3, low2, EFEH, Moghe, Linear, patent, hybrid}. EFEH is first proposed for high and middle voltage overhead power lines by regarding the E-Field in abundance. It is then applied to low voltage systems as mounting single-phase AC power cords with metallic sheaths \cite{low3, low2, EFEH}. These efforts revealed that, it is also possible to constitute an applicable EFEH methodology for applications in which E-Field intensity is considerably low. As an alternative to the previous studies, Linear Technology (LT) has brought a new perspective to the area with their parallel plates model \cite{Linear}. This work includes placing copper plates under the fluorescent troffers to exploit the surrounding field. The plates act as a capacitive voltage divider, block the outward field flow, and utilize the leakage electric charges to provide stable voltage. The idea presented by LT is taken as the basis of this paper, and modified to build a more flexible and efficient EFEH model. 

\section{Principle}

\begin{figure}[!t]
\centering
\subfigure[]{
\includegraphics[width=0.275\textwidth]{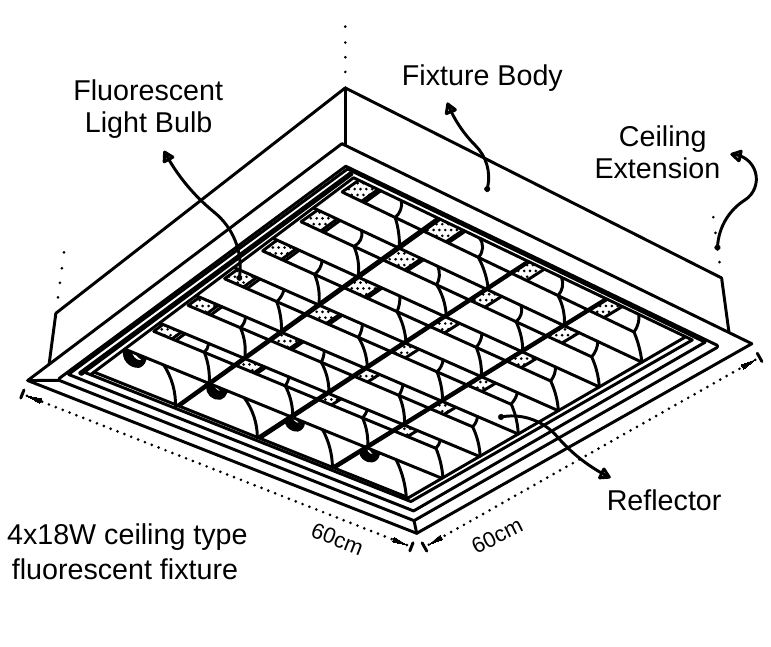}}
\subfigure[]
{\includegraphics[width=0.195\textwidth]{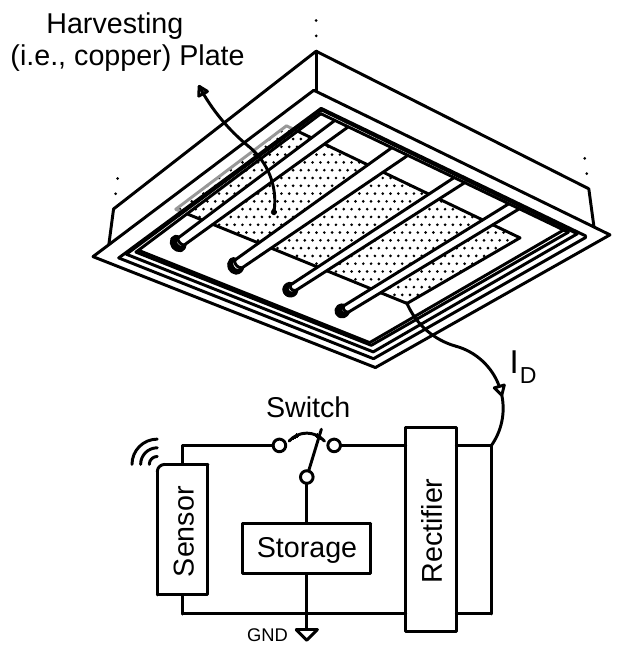}}
\vspace{-3mm}
\caption{Depiction of the harvester system 
\source{\textit{a}~~Conventional overhead $4$-light fluorescent troffer/fixture model}
\source{\textit{b}~~The proposed EFEH concept (patent pending design \cite{patent})
}}
\label{fig:1}
\end{figure}

\begin{figure}[!b]
	\subfigure[]{
		\includegraphics[width=0.33\textwidth]{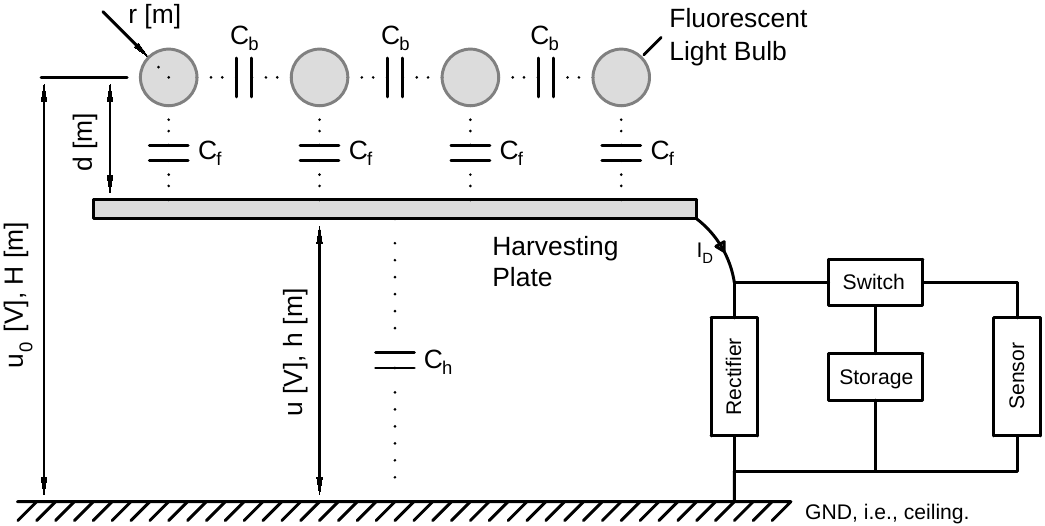}}
	\hspace{0.5mm}
		\subfigure[]{
		\includegraphics[width=0.13\textwidth]{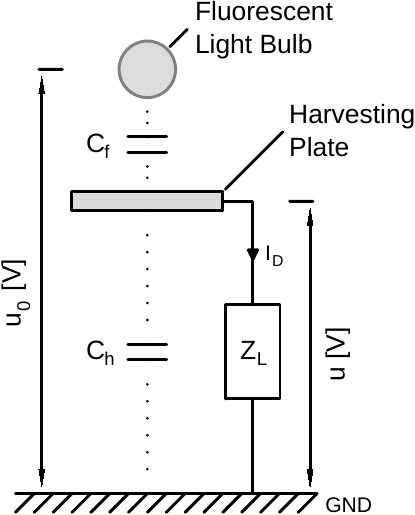}}
	\vspace{-3mm}
	\caption{Distribution of the formed capacitors
		\source{\textit{a}~~Physical model of the proposed EFEH concept}
		\source{\textit{b}~~Simplified model of the proposed EFEH concept}}
	\label{fig:2}
\end{figure}

Fig.~\ref{fig:1} roughly depicts the basics of our proposed E-field energy harvesting concept. As shown in Fig.~\ref{fig:1}(b), a copper plate, i.e., capacitive voltage divider, is situated between ceiling and the field emitting fluorescent light bulbs. Splitting up the field by a conductive material not only results in a voltage difference, but also formation of stray in-plane capacitances as in Fig.~\ref{fig:2}(a).
In this figure, capacities from bulbs to harvester are stated as $C_{f}$ while the residual ones, between the harvester and ground planes, are titled as $C_{h}$. It is also required to mention of stray parasitic capacitors, $C_{b}$, formed between the fluorescent bulbs due to oscillating line voltage. The electric charges collected on the harvester are transferred by $I_{D}$, and accumulated in $C_{s}$ after rectified. The given `Switch' model refers to an autonomous power conditioning (APC) circuit to be employed for switching between harvesting and transmission stages.

With the given model, Fig.~\ref{fig:2}(a), it becomes possible to estimate the available power that can be extracted from the ambient field. Considering the measurements we took with an electromagnetic radiation tester, fluorescent tubes are assumed as a uniform source of E-field in order to simplify the system model, and therefore reduce the computational complexity. Under this circumstance, the major contributions are only due to one $C_{f}$, and a serially connected $C_{h}$ as illustrated in Fig.~\ref{fig:2}(b). The capacitors, $C_{f}$ and $C_{h}$, can be stated as \\
\begin{equation}
C_{f} = \dfrac{2\pi \epsilon l}{\cosh^{-1}(d/a)}~~~\textup{\&}~~~C_{h} = \dfrac{2\pi \epsilon l}{\ln(2H/a)}~~\textup{[F]}~~ 
\label{eq:1}
\end{equation}
where $a$ is the tube radius, $l$ is the tube length, $d$ refers to vertical aperture between the tube and the harvester plate, and $H$ denotes the distance from tube's center to ground.

Regarding the resultant voltage divider formed by simplified model the equivalent impedance $Z$ and the load voltage $u$ can be stated as
\begin{equation}
Z = \dfrac{Z_{L}}{1+jwC_{h}Z_{L}}+\dfrac{1}{jwC_{f}}~~~\textup{\&}~~~u = \bigg(\dfrac{Z_{L}}{1+jwC_{h}Z_{L}}/Z\bigg).u_{0}
\label{eq:2}
\end{equation}
where $u_{0}$ and $w$ represent rms voltage and angular frequency of the power-line, respectively. 
If the expressions in (\ref{eq:2}) are combined, the load voltage can be expressed in terms of load impedance $Z_{L}$ as
\begin{equation}
u = \dfrac{jwC_{f}-w^2C_{f}C_{h}Z_{L}}{jw(C_{f}+C_{h}(1+Z_{L}))-w^2C_{h}Z_{L}(C_{h}+C_{f})}.u_{0}
\label{eq:3}
\end{equation}

If $Z_{L}$ is assumed as an ohmic load, the obtainable power from the harvester can be calculated as
\begin{equation}
P = \dfrac{|u|^2}{Z_{L}}~~\textup{[W]}~~
\label{eq:4} 
\end{equation}

When $\partial P$/$\partial Z_{L}$ = $0$, the optimal load impedance $Z_{L_{opt}}$ is obtained. The maximum attainable power is therefore expressed as $P_{max}$ = $P(Z_{L_{opt}})$. By regarding this theoretical investigation and the dimensional limits of the fluorescent fixture to be employed, the size; position; and the structure of the E-field energy harvester can be determined to obtain the best performance achievable.

\begin{figure}[!t]
	\subfigure[]{
		\hspace{-5mm}
		\includegraphics[width=0.32\textwidth]{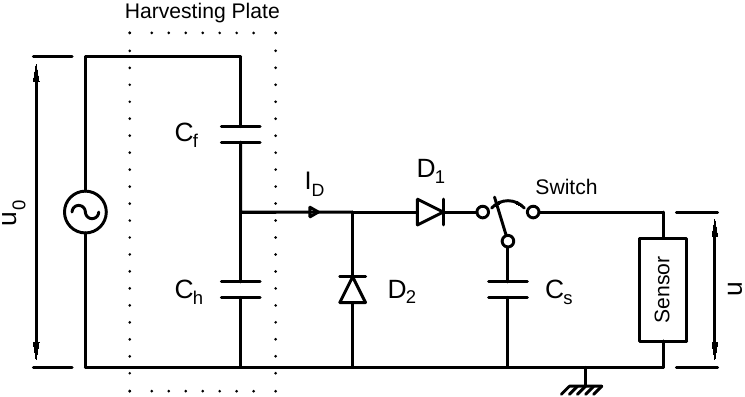}}
	\subfigure[]{
		\hspace{3mm}
		\includegraphics[width=0.15\textwidth]{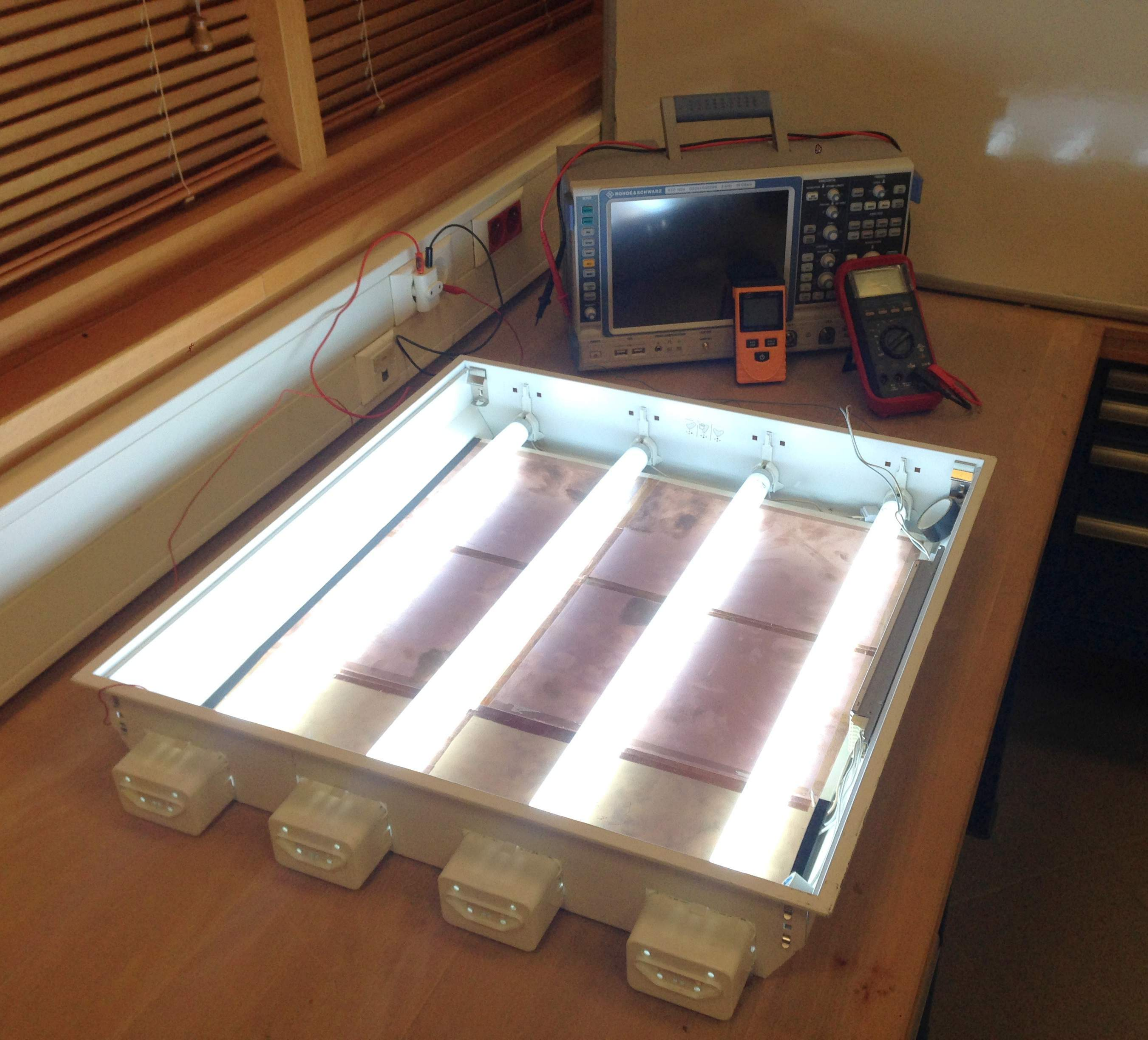}}
	\vspace{-3mm}
	\caption{EFEH system schematic, and test-bed
		\source{\textit{a}~~Equivalent circuit of the proposed EFEH concept}
		\source{\textit{b}~~Experimental Setup}}
	\label{fig:3}
\end{figure}

\section{Performance Evaluation}

In order to validate the competence of our proposal, a conventional $4$x$18$W-T8 type $60$x$60$x$10$cm in dimension overhead fluorescent troffer is utilized. We situated a copper plate above the fluorescent bulbs to obstruct the field that is being emitted. The plate, i.e., the harvester, leaks the electric charges, and a rectifier, $D_{1}$ and $D_{2}$,  converts them into DC as minimizing the switching time while preventing back feeding. A quick-charged; long-lasting; and high power condensed $0.1$F-$5.5$V super-capacitor, $C_{s}$, is employed to store the converted energy. The interchange between harvesting and nodal operation modes is performed by the APC, which is simply constituted by a voltage comparator and a corresponding FET. This circuit autonomously enables charge transfer when the accumulated energy is high enough to power the sensor, and switches back to harvesting stage when the voltage on $C_{s}$ descends below a predefined threshold. This action not only prevents the discharge of $C_{s}$ to $0$V, but also allows more frequent transmissions by shortening the time exerted on power extraction \cite{EFEH}. A detailed depiction of the above-mentioned design is illustrated in Fig.~\ref{fig:3}(a).

\begin{figure}[!b]
	\centering
	\subfigure[]{\includegraphics[width=0.225\textwidth]{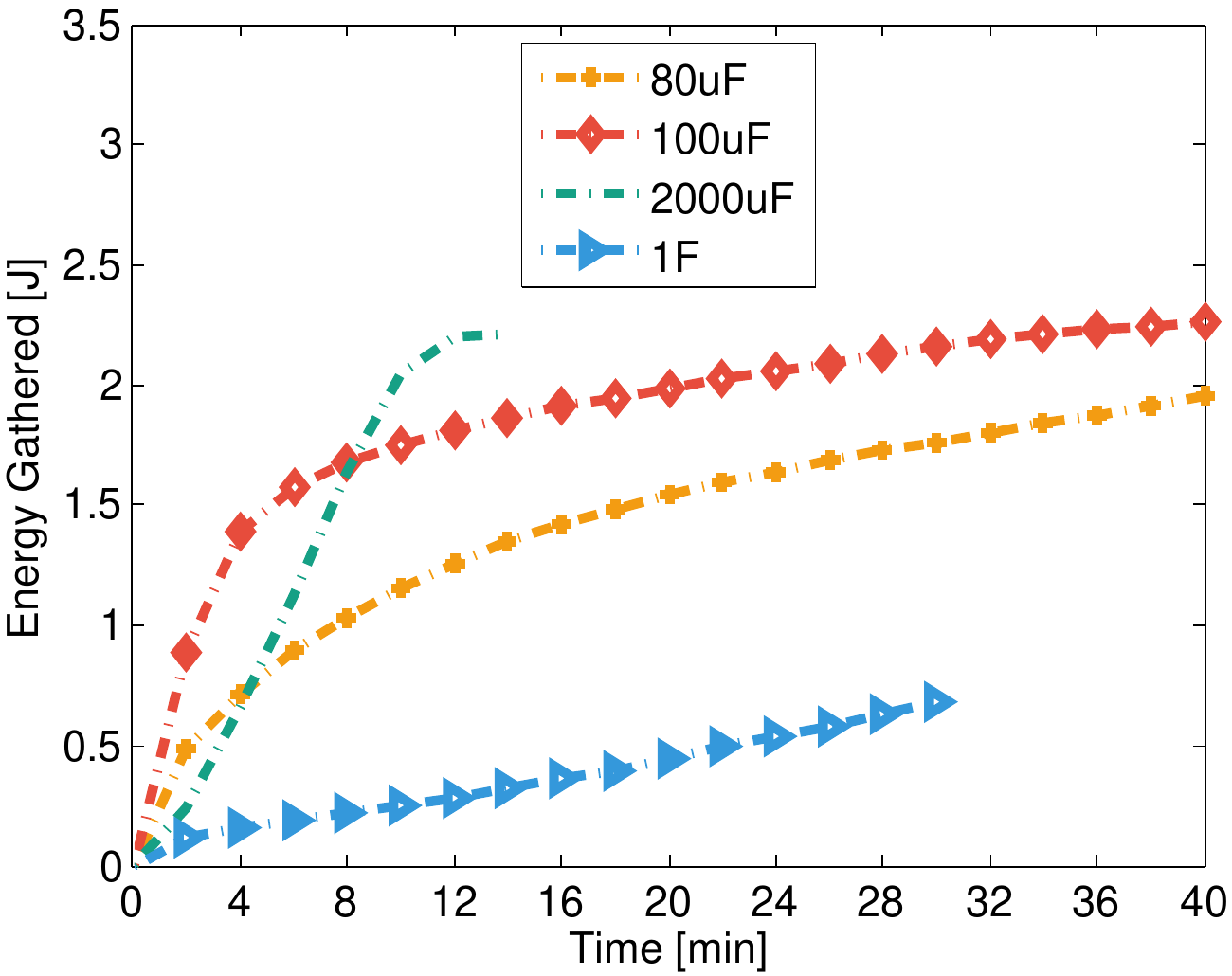}}
	\hspace{3mm}
	\subfigure[]{\includegraphics[width=0.225\textwidth]{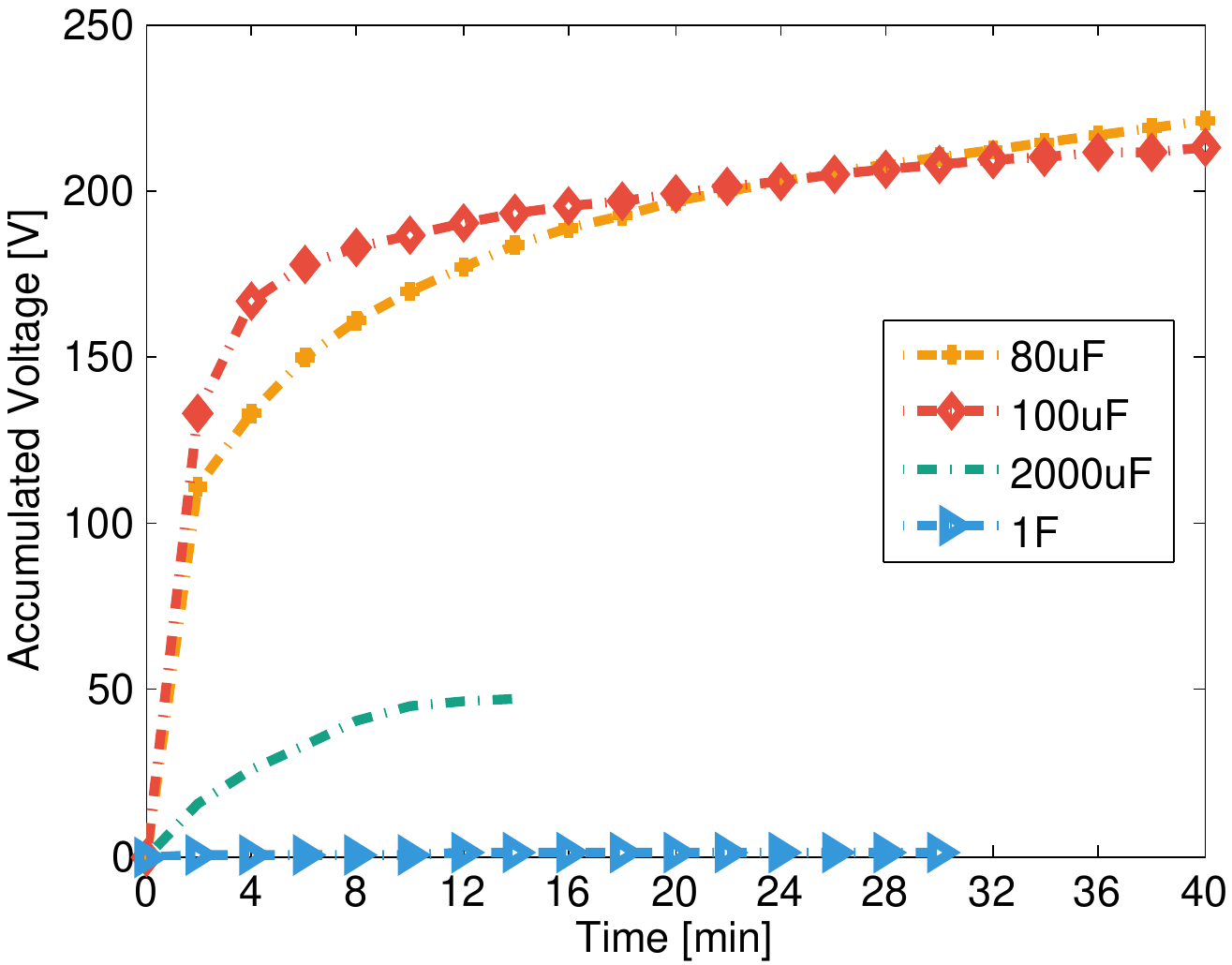}}
	\vspace{-3mm}
	\caption{Experiment results -for electronic circuitry optimization  
		\source{\textit{a}~~time vs. accumulated voltage patterns for various capacitors}
		\source{\textit{b}~~time vs. gathered energy patterns for various capacitors}
	}
	\label{fig:4}
\end{figure}

\begin{figure}[!t]
\centering
\subfigure[]{\includegraphics[width=0.225\textwidth]{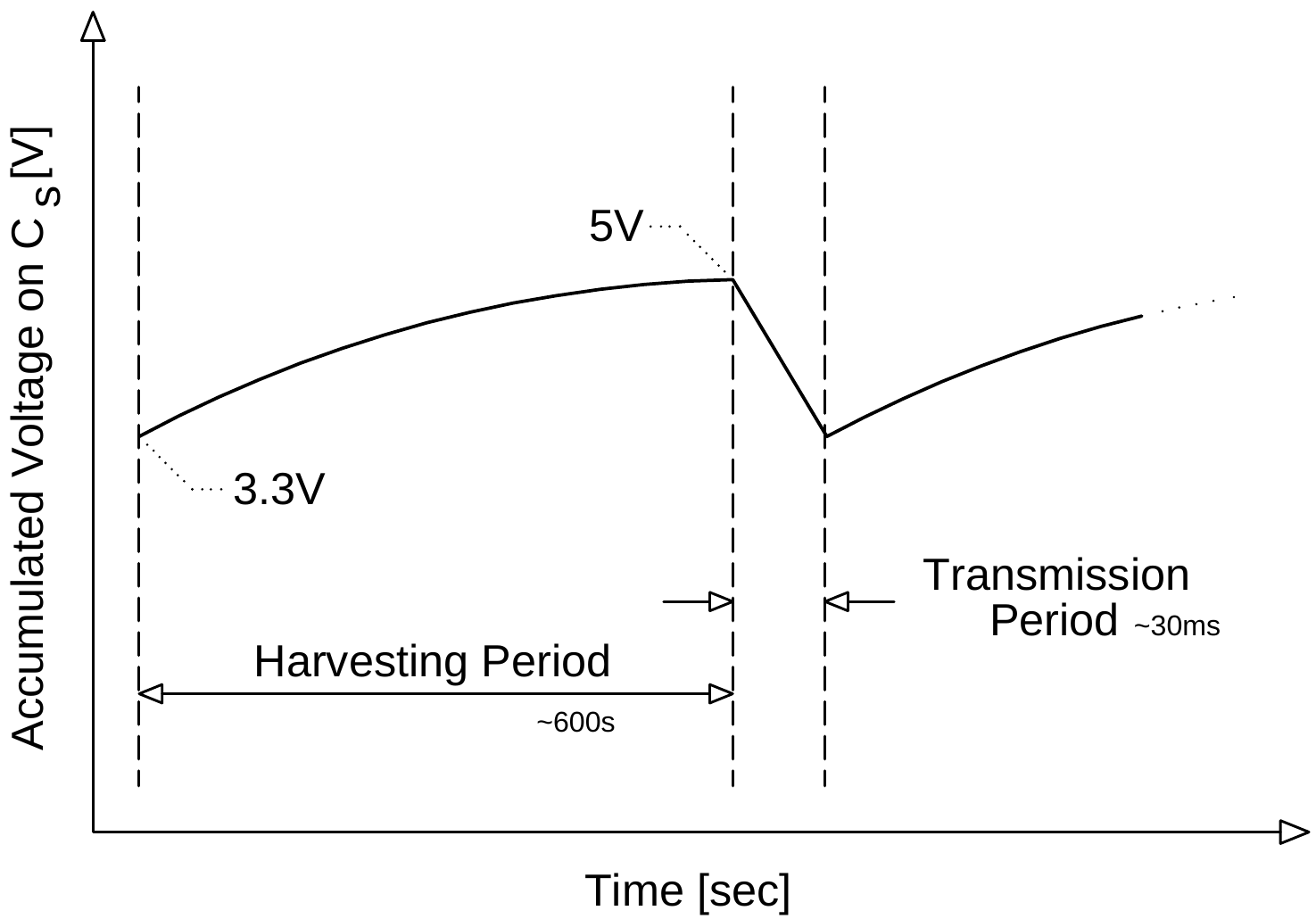}}
\hspace{3mm}
\subfigure[]{\includegraphics[width=0.225\textwidth]{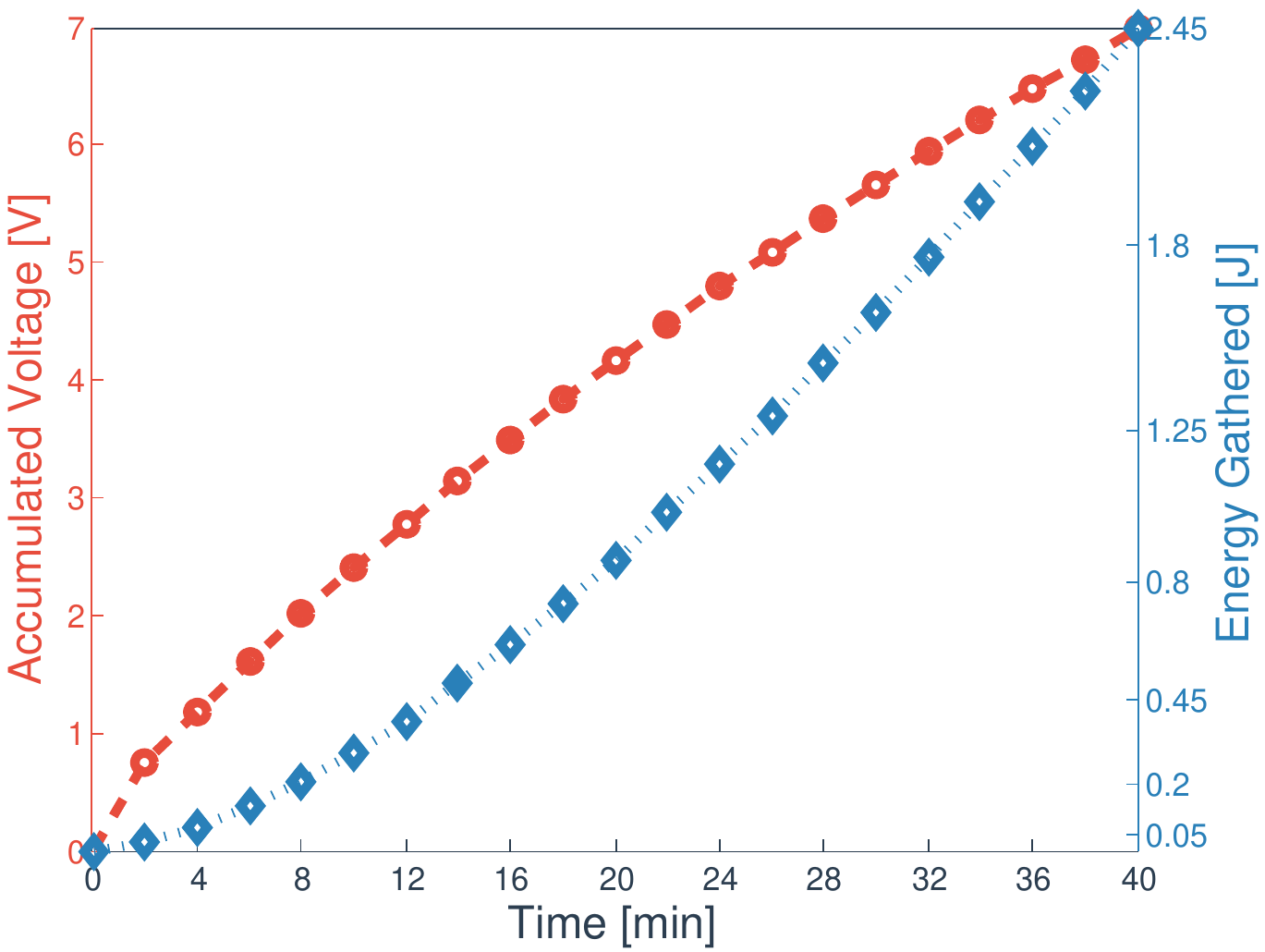}}
\vspace{-3mm}
\caption{Finalized circuitry outcomes 
\source{\textit{a}~~Duty cycle depiction of the proposed EFEH concept}
\source{\textit{b}~~time vs. accumulated voltage/energy on $C_{s}$}
}
\label{fig:5}
\end{figure}

We took several measurements on a conventional fluorescent troffer, Fig.~\ref{fig:3}(b), for both estimating the energy density, and the optimal circuitry that needs to be structured. In this direction, we utilized various types of storage elements, and analyzed their performance to compromise between charging time and the accumulated voltage, Fig.~\ref{fig:4}(a). The results in Fig.~\ref{fig:4}(b) reveals that, it is possible to scavenge $1.25$J of energy, on average, in $25$min with our proposed model. When the basics of electrostatics is considered, $0.1$F super-capacitor stands as the best option as saturating at roughly $5$V which is the exact voltage required for the nodal operation. Thus, it becomes unnecessary to employ any buck and/or boost converter-like component to further adjust the output voltage. As the constraints related to circuit complexity are resolved, power transfer efficiency, i.e., the level of power that is transferred to the resting parts of harvesting system, is also enhanced thanks to this approach. As a result, $3.3$V level is set for the harvesting threshold, and the resultant pattern of the APC, i.e., duty cycle of the EFEH system, is illustrated in Fig.~\ref{fig:5}(a). 
After waiting sufficient time, it is about $2.45$J of energy was accumulated in $C_{s}$ as seen in Fig.~\ref{fig:5}(b). These results refer to a remarkable improvement in harvesting efficiency when the existing power provision architectures are considered.

The proposed harvesting procedure has the benefit of interruption-free employment due to the presence of neutral bus situated near-by. In other words, there is no need for peeling of concrete surface to complete the harvesting circuit on the contrary of previous studies \cite{low3, low2}. This model also yields in more secure implementation by requiring no galvanic contact with the field emitting assets unlike magnetic field-based counterparts \cite{Moghe}. We therefore obtain a unique solution that offers flexible, safe and cost efficient installation. In addition to these outstanding features, the proposed configuration stands as an interdisciplinary effort, which opens up the potential of a hybrid harvesting technology \cite{hybrid}. Since some part of the energy is dissipated as heat during illumination, the harvester can be structured as enabling power extraction from temperature gradients in addition to E-field. Furthermore, the lights emitted from fluorescent tubes can be also exploited if the harvester is structured with capability of PV conversion.

In addition, to not affect the luminaire efficiency due to lower reflectivity of the copper plate as against matt-white floor of the fixture chest, we altered our model by mounting the harvester with a more reflective white paper. The measurements taken by a lux meter point out an increment in illimunance by $5.2\%$. This enhancement shows that our proposal not only operates without affecting the illumination process but also offers increased luminaire efficiency on the contrary of \cite{Linear}. The proposed architecture therefore stands as a promising candidate for broadening the scope of energy harvesting communications, and to ease the building of battery-less IoT networks by its eligible aspects. 

\section{Conclusion}

This paper presents a new power provision methodology that exploits the outward E-field flow around the illumination assets. The most prominent features being offered by this work can be referred as reducing the circuit complexity, increasing the luminaire efficiency, and providing abundance of energy with minimal installation, design, and maintenance costs. Empirical results revealed the potential of this method for the applications in which greater longevity; higher robustness; and larger throughput is crucial. It is believed that our approach will alleviate the energy constrained IoT services in the very near future by substituting the batteries.

\vskip5pt

\noindent O. Cetinkaya (\textit{Department of Electrical and Electronics Engineering, Koc University, Istanbul 34450, Turkey})
\vskip3pt

\noindent E-mail: okcetinkaya13@ku.edu.tr
\vskip3pt
\noindent Ozgur B. Akan (\textit{Department of Engineering, Electrical Engineering Division, University of Cambridge, Cambridge CB2 1TN, UK})

\end{document}